# Insights into Transient Dynamics of Bacteria Laden Liquid Bridges

*Kush Kumar Dewangan[1], Srinivas Rao S[1‡], Durbar Roy[1‡], Atish Roy Chowdhury[2],*

*Dipshikha Chakravortty[2,4] , Saptarshi Basu[1,3]\**

[1]Department of Mechanical Engineering, Indian Institute of Science, Bangalore 560012, Karnataka, India
[2]Department of Microbiology & Cell Biology, Indian Institute of Science, Bangalore 560012, Karnataka, India
[3]Interdisciplinary Centre for Energy Research (ICER), Indian Institute of Science, Bangalore 560012, Karnataka, India
[4]School of Biology, Indian Institute of Science Education and Research, Thiruvananthapuram Karnataka, India
\*Corresponding author email:sbasu@iisc.ac.in

‡ S.R.S., D.R. contributed equally to the work.

**ABSTRACT**

We study the evaporation mechanics and precipitate characterization of bacteria laden liquid bridge using experimental and theoretical analysis. Aqueous suspension of motile and non-motile Salmonella Typhimurium and Pseudomonas aeruginosa typically found in contaminated food and water were used in liquid bridge configuration between hydrophilic substrates. Using the inverse logarithmic evaporation flux model, we study the liquid bridge volume regression using cylindrical/catenoid volume models for various confinement distances. We show that for all confinement distances, the volume regression is linear on normalizing both volume and time as in the case of a pure sessile drop. However, in normalized volume and dimensional time space we observe some non-linearities as the total evaporation time scales non linearly with the confinement distance. The non-linearities found were captured using the catenoid model. The cylindrical volume model fails to capture the regression trend whereas, the catenoid model conforms to the experimental volume regression data at all confinement distances, and the transient liquid bridge interface evolution profile at high confinement distance. Further, we also study the precipitate pattern and bacterial distribution using micro/nano characterization techniques. We show the average precipitate pattern for both sessile and higher confinement distance resembles coffee ring type deposits although the underlying bacterial distribution differs. For lower confinement, we observe pattern resulting from a combination of coffee ring effect, stick-slick motion, and thin film instability. The reduction in confinement distance causes an altered bacterial agglomeration, resulting in a multi-pattern network instead of a single circumferential edge deposition. We show the aerial size of motile bacteria increases with decreasing confinement, whereas the size for non-motile bacteria remains constant in the precipitate. These findings have a crucial understanding on the establishment of pathogenesis in humans in the hospital settings.

KEYWORDS: *Salmonella* Typhimurium (STM), *Pseudomonas aeruginosa*, Liquid bridge (LB), Atomic Force Microscopy (AFM), diffusion-driven model.

1. **INTRODUCTION**

A liquid bridge is a fluid configuration where a mass of liquid is sustained between two solid surfaces or liquids. Liquid bridge (LB) configuration has been extensively studied during the past few decades because of its importance in various medical problems such as respiratory diseases [1,2], the behavior of body joints, the feeding of shore birds [3], and the food canal of monarch butterflies [4]. This distinct phenomenon also appears in many industrial processes, such as printing processes [5], materials engineering [6], oil refineries [7], and flow in porous media [8]. Moreover, LB can be found between soil particles and the root surface of the plant [9].

The evaporation dynamics of LBs with suspended materials confined between solid surfaces have been analyzed extensively [10–13]. It differs from the evaporation of a sessile drop primarily in three aspects: contact angle, the height of the bridge, and capillary force [10]. However, the major difference between LB and sessile droplet evaporation originates due to the dynamics of the drastically different geometry and the associated capillary force. The deposition pattern formed after the disengagement and evaporation of the colloidal LB may be of greater interest to several industrial and biological processes. Basic experimental studies have been performed to understand how different patterns are formed during the drying process of LB on various surfaces [12–14]. Drying of LBs exhibits a rich phenomenology that depends on the suspended materials (physical properties, shapes, etc.), convection and evaporation phenomena, surface tension and capillary interactions, contact line pinning and depinning, membrane stretching and bending (when LB formed between membrane and solid surface), Marangoni forces and hydrophobicity. Drying patterns of the colloidal drop may also depend on many factors such as suspension particle shape [15] as well as size [12], surface wettability [16], surfactant concentration [17], ambient pressure [18], surface temperature [19] and type of solutions. Generally, the pinning and depinning (stick-slip motion) at the contact line governs the deposition patterns. The contact line can stay pinned to form a single ring (i.e., "coffee ring") or can shrink in a discontinuous manner to generate multiple rings. Patil et al.[20] categorized the deposition pattern into three types, namely, a ring, a thin ring with an inner deposit, and an inner deposit. Recently, Yang et al.[21] reviewed different deposition patterns of drying droplets: (a) coffee-ring; (b) mountain-like; (c) volcano-like; (d) multi-ring; (e) eye-like; (f) spoke-like. Some other patterns, such as ribbon-like [15], Scallop Shell like [14], and spider web-like [22], are also reported.

Researchers from different backgrounds have investigated the evaporation dynamics of a sessile drop laden with bacteria and the subsequent emergent patterns due to solvent depletion [23–26]. However, limited attention is paid towards the drying droplets in the capillary LB configuration, particularly in the case of an LB populated with bacteria. In many biological processes, this type of confined drying of liquid plugs is observed, particularly in respiratory systems [27]. In general, coffee-ring effect is used for disease diagnosis with the deposited bacteria along the contact line during evaporation [28]. The deposited bacteria at the contact line are severely affected by the local interactions and collective dynamics, which can help them survive and spread over surfaces. Deposited patterns of bacteria also depend on motile and non-motile nature [29]. The drying pattern in the horizontal open-ended capillary tube (length 2.2 cm, inner diameter 0.15 cm) under ambient conditions with active matter (tobacco mosaic virus) was reported by Lin et al.[30]. They observed that stick–slip motion forms the periodic pattern on the surface. Thokchom et al.[31] experimentally studied the effect of chemotaxis on the deposited pattern of live and dead bacterial suspensions drop. Andac et al.[32] presented bacterial mobility and its significant role in slow evaporation.

Liquid bridge evaporation on a substrate is a complex process involving several factors including evaporation, hydrodynamics, wetting/dewetting dynamics, etc. The diffusion-controlled evaporation model is adopted in the literature [10,12,33–36] to describe the evaporation characteristic of a steady LB by neglecting the external convection. Upadhyay & Bhardwaj[12] developed an evaporation model based on the Laplace equation for the liquid−vapor concentration. In most diffusion approaches assumed a cylinder for concave LB, and the results agreed with experimental data of Upadhyay & Bhardwaj[12]. However, evaporation regression analysis with a catenoid-shaped LB (cosine hyperbolic function) has not been previously reported. The surface of the LB also contributes to the evaporation rate. The role of surface wetness on the evaporation dynamics of LB is examined by Harimi et al.[33].

The LB evaporation is typically slow under normal ambient conditions. The LB contact line detaches from surface, and as evaporation takes place, the system adjusts to a new quasi-stable state. Delaunay[37] first developed analytical methods to analyze the meridional profile of liquid bridges of revolution using surfaces of revolution with constant mean curvature. According to this, the meridional profile of the liquid bridge could be a part of a nodoid, unduloid, or, in some cases, a catenoid, cylinder, or torus. The geometry of a steady LB surface is described by the Young–Laplace equation, according to which the mean curvature of the surface is constant in the absence of gravity [38,39]. Harimi et al.[33] coupled the diffusive flux model with the Young-Laplace equation to evaluate the volume, shape, and capillary pressure of LBs as evaporation proceeds. Wang et al.[38] analytically studied the shape of symmetric and asymmetric LB. They have verified the analytical solution for different fluid drops (water, Kaydol, and dodecane) for 1 $\mu$l volume of LB with height 0.91 mm. A transient LB interface evolution profile using the evaporation model has not been reported in the literature. On the other hand, Ban & Son[40] numerically studied the two-dimensional flow and thermal characteristics of LB evaporation using the level-set method. They suggested that one-dimensional analytical solution is not suitable for the higher rate of evaporation.

In this sequence, the energy minimization method [39,41] is the most effective approach to predicting the profile of LB. The equilibrium profile of the liquid–vapor interface can be determined from the curve associated with the minimum potential. Teixeira and Teixeira[42] calculated the conditions for the existence and positions of any necks or bulges and inflection points on the surface of the LB.

Previously several researchers studied the LB not only with conventional fluid but also with bacterial solution [43]. It is possible to understand biofilms better by studying bacterial bridges, the structures, and functions of complex 3D biofilms, as well as factors that can affect biofilm formation and the removal of biofilms with the assistance of bacterial bridges. The bio-microfluidic channels are used to study bacterial bridges between bacterial colonies of PAO1 [43]. There is a possibility that these biofilm bridges are common in nature. PAO1, E. coli, and *Salmonella* Typhimurium (STM) are the most common bacteria in the wastewater.

*Salmonella* Typhimurium (STM) causes food-borne diseases, including gastroenteritis, diarrhea, and typhoid fever in humans frequently, and their infection continues to pose a severe health threat in countries throughout Asia and Africa [44–47]. Available literature suggests that *Salmonella* Group B spp. is the primary cause of lung abscess in humans. Pitiriga et al.[48] reported one of the first cases of a lung abscess caused by *Salmonella* serovar Abony in an immunocompetent healthy young adult. In the elderly, *Salmonella* infection is responsible for the majority of cases of aortic and other vascular infections [49,50]. Another Gram-negative pathogen *Pseudomonas aeruginosa* causes chronic lung infection in patients suffering from cystic fibrosis [51].

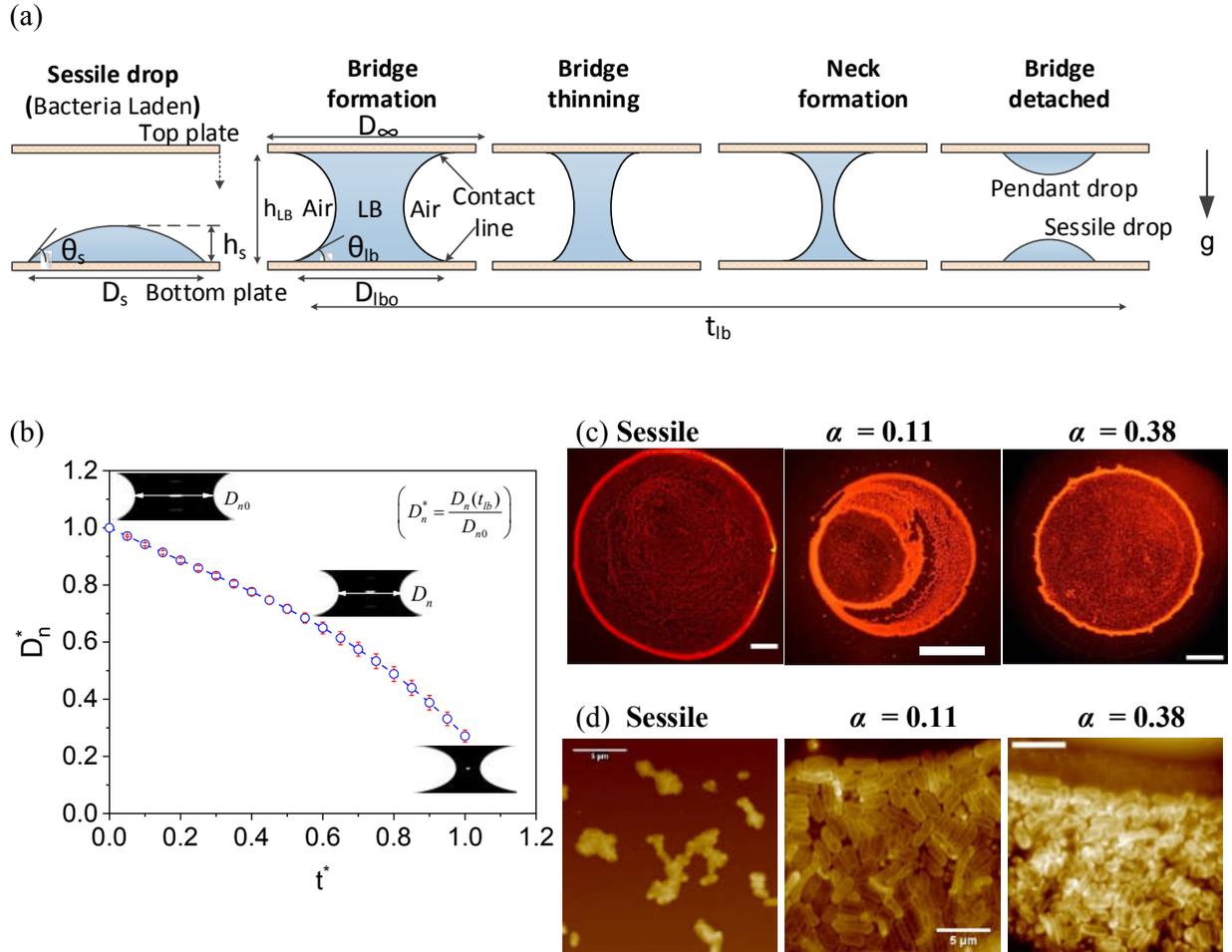

**Figure 1. (a)** Schematics chronology of the capillary LB formation and its life times with defining different terms. The arrow with a dashed line denotes the movement for forming LB. **(b)** A typical temporal variation of non-dimensional neck diameter $\left(D_n^* = D_n(t_{lb})/D_{n0}\right)$ with time $t_{lb}^*$ (normalized against the individual $t_{lb}$) for the case of non-dimensional confinement distance ($\alpha = h_{LB}/D_{lb0}$) 0.38 for milli-Q. **(c)** final deposition on the bottom surface (The scale bar is 0.5 mm) and **(d)** AFM image of sessile and LB ($\alpha = 0.11, 0.38$) (The scale bar is 5 μm).

As bacteria-laden droplets are released from contaminated water sources or during the respiratory activities of an infected person (coughing or sneezing), they typically fall into many substrates (fomites). When such a fallen droplet is in direct contact with the other surface or hand of the human, it can make an LB. A brief review of the literature reveals that the studies available are limited to identifying the bacterial deposition formed by the evaporation of LB confined between parallel plates. The phenomenon of bacterial LBs helps in a better understanding of biofilms during the evaporation and physical deposition mechanisms of bacteria that yield a specific pattern. A multiscale comprehensive approach is presented to understand the evaporation induced bacterial deposition and pattern formation in confined evaporating LB between two glass surfaces at variable separation distances. Sessile droplets placed on the bottom substrate were positioned parallel to the top substrate to rearrange it into the LB, and variations in the evaporation dynamics at three confinement distances during the evaporation were considered (refer to Fig.1). The evaporation characteristics of LB are discussed with the help of an analytical solution based on the diffusion at the liquid-vapor interface. In the present work, the analytical method of Upadhyay and Bhardwaj[12] is extended to the case of a hyperbolic curve of an LB with midplane symmetry. We investigated the effect of the evaporation modes (LB and sessile) on the deposited patterns. The influence of the gaps between the two plates on the deposited patterns and bacterial morphology is discussed. Together with these, the present work also covers the effect of live and dead bacteria on the deposited patterns.

## 2. MATERIALS AND METHODS

### 2.1. Preparation of bacterial suspension

The bacterial sample is prepare using single colonies of wild-type *Salmonella* Typhimurium (STM) and *Pseudomonas aeruginosa* (PAO1) were inoculated in 5 mL of LB broth and incubated overnight in a shaking incubator. After 12 hours of incubation, stationary phase bacterial cultures were centrifuged at 6000 rpm for 10 minutes, and the pellet was washed with sterile milli-Q water. Finally, the pellets were resuspended in 1.5 mL of milli-Q solution (Bacterial colony forming unit, CFU~ 1.5 to $3\times10^8$). For microscopic imaging, the bacterial cells were stained with 1 µg/mL of propidium iodide.

### 2.2. Experimental setup for LB dynamics & fluorescence imaging

In the present study, plain glass slides of dimensions 25×25×1 (length × breadth × height) $mm^3$ are used to construct a parallel plate confinement assembly for the drying experiments. The square standard glass was sliced into square-shape (purchased from Blue Star©) with a glass cutter to obtain the required equivalent dimension. In order to minimize the edge effects, square plain slides are used. Prior to experiments, the glass slides were kept in an isopropanol bath for 15 min, followed by cleaning with Kimwipes (Kimberly Clark International). Temperature and relative humidity were controlled at 25±3°C and 45% ±3%, respectively, throughout all experiments (measured with a sensor TSP-01, provided by Thorlabs). A droplet volume of 0.4 ± 0.1 µl was gently placed onto the central part of the bottom substrate (glass slide) using a micropipette (procured from Thermo Scientific Finnpipette, range: 0.2–2 µl), and later, a micrometer was used to position the top slide accordingly to come in contact with the bacteria-laden droplet. Afterward, the drop was allowed to dry in a liquid bridge (LB) configuration.

A digital camera (Nikon D5600) fitted with a 2X zoom lens assembly (Navitar) was used to record the evaporation dynamics of the LB at 30 frames per second from the side view. LB is illuminated from the light Source (50 - 250 W Quartz Tungsten Halogen Research Light Sources), as shown in Figure 2b.

Deposited patterns of the LB on the glass slides were acquired from the top using an optical microscope (Olympus). The light source in line with the objective lens (high magnifications 10× & 20×) of the microscope illuminated the drop, and the CCD camera (Nikon D7200), mounted on top of the microscope, captured the images of the deposit (Figure 2b).

### 2.3. Interferometry image analysis

The receding phenomena of the microlayer for various heights of capillary bridges have been captured by employing interference microscopy. Microscopic alignment and the incorporated components are shown in Figure 2b. A pulse-laser source of 640 nm wavelength (Cavitar, Finland) of a light source is positioned on a beam splitter that transmits and reflects the light beam. The reflected light beam passes through the microscopic objective (4X and 20X, LYNX) and reflects back from the bottom glass plate surface positioned at the focal length of the microscope. The developed phase difference of the light beam at the leading edge of the capillary bridge and the glass surface results in the measurement of the film thickness. It passes through the objective, beam splitter, zoom lens (Navitar), di-choric mirror, as seen in Figure 2b. A high-speed camera (Photron SA-5, Lavision) is used to capture the interference fringe patterns caused by the phase difference change. The data reduction process and the image analysis for obtaining the two-dimensional distribution of the sequential steps have been reported in detail by Chattopadhyay et al.[14], Rasheed et al.[52], and Roy et al.[53].

**Atomic Force Microscopy**

An atomic force microscopy (AFM) system (Park System, South Korea) combined with a flat scanner and an optical microscope. For different samples, the scanner is utilised to gather data on adhesion energy, variation of force-distance spectroscopy, and microscopic scanned pictures in both contact and non-contact modes. The images were acquired at 512 lines/scan at 0.8–0.9 Hz scan rate. The XEI software is used to analyze raw AFM data obtained from the bacterial sample. AFM in contact mode is performed using CONTSCR (256 pixels, 0.5 Hz scan rate) with a scanning radius of less than 8 nm. In order to acquire the force indention curve, the stiffness of the cantilevers was 0.2 N/m with a resonance frequency of 25 kHz. The cantilever sensitivity and spring constant are calibrated before every experiment.

## 3. RESULTS AND DISCUSSIONS

### 3.1. Formation of the liquid bridge and evaporation dynamics

The primary objective of this study is to investigate the evaporation dynamics and deposits caused by bacterial agglomeration in an evaporating LB under different confinement distances. The variation in the contact line (CL) and curvature of the LB during the evaporation period is recorded using a digital camera. The LB is created between two plain glass slides schematically shown in Figure 1(a). Figure 2 depicts the experimental setup used to measure evaporation dynamics, and thin film thickness. STM wild-type (WT) and *Pseudomonas aeruginosa* (PAO1), live as well as dead bacteria, were incorporated in milli-Q solution to understand the evaporation dynamics, and deposition patterns. The bacterial-laden LB is referred to as LB in this work.

In order to provide a better understanding of the deposit patterns, we will first discuss briefly the temporal events that occur during the lifespan of an LB, beginning with its creation and ending with its eventual evaporation.

A colloidal sessile drop was placed symmetrically at the center of the bottom plate, and the top plate was positioned parallel to the bottom plate. With the help of a micrometer, the top plate is adjusted to touch the sessile drop, and the sessile drop rearranges itself into the LB configuration (refer to Figure 1a). The sessile drop contact diameter (CD), contact angle (CA), and droplet height are denoted as $D_s$, $\theta_s$, and $h_s$, respectively. A confinement distance between two parallel slides is expressed by $h_{LB}$. The complete evaporation time of LB is denoted by $t_{lb}$. The CD, neck diameter, and CA in the LB configuration are denoted by $D_{lb}$, $D_n$, and $\theta_{lb}$, respectively. Subscript 0 represents the initial state of the LB. The transmission of fluid from the bottom plate to the top plate surface reduces the CD at the bottom plate as the drop attempts to conserve its volume (refer to Supplementary Video-1). The LB height is determined to be the confinement distance, and the bridge is now allowed to dry. As time progresses, the LB evaporates; the solvent volume gradually reduces, and finally LB neck breakup occurs. The LB formation and lifetime can be divided into five different stages: sessile drop, stable LB formation, LB thinning, thin neck formation, and detachment of bridge (refer to Figure 1a). The neck diameter reduction of the LB with evaporation is plotted in Figure 1c, three stages seen here. It is observed that after detachment of the LB, individual drops form on both plates (see the final image in Figure 1a), and they are dried upside down as a pendant drop on the top surface is indistinguishable from a sessile drop on the bottom.

Though the LB and sessile drop evaporations have identical evaporation conditions (RH, temperature, particle concentration, volume, etc.), the internal flows are significantly different [14]. For the LB, bacteria migrated towards CL by the outward capillary flow, and the compelling

Marangoni flows are also driven by surface tension gradients [13]. Two factors contribute to the development of the surface tension gradient: (1) the evaporative flux is higher at the three-phase CL, which results in a temperature gradient, and (2) a gradient in the concentration of solute between the three-phase contact line and to the center of the neck (liquid-vapour interface). Evaporation flux depends on the surface's curvature [12] and influences the flow inside LB [14]. With an increase in the confinement spacing, the radius of curvature of the liquid meniscus and the three-phase contact angle decreases [14], and the liquid–solid contact area decreases.

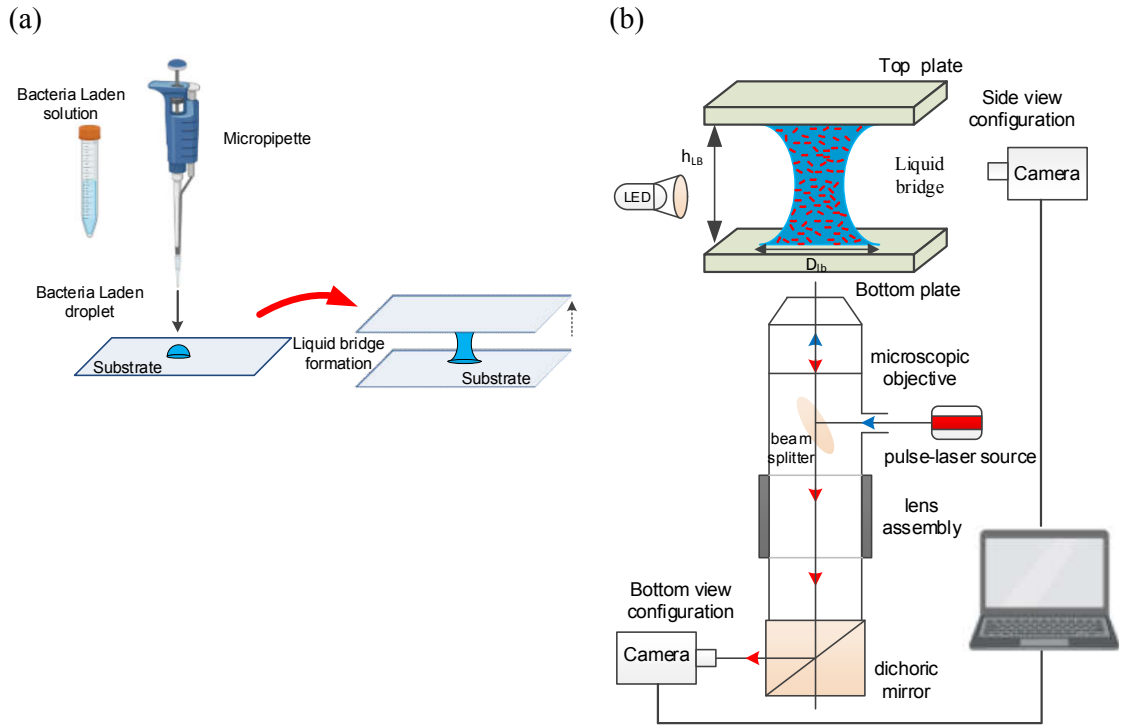

**Figure 2.** (a) Schematically represents the creation of a liquid bridge from a sessile drop.
(b) The schematic representation to capture the evaporation dynamics backlight imaging and microscopic interferometry setup.

Figure 1(a) shows that pendant and sessile drops form after the snapping of the LB. Typically, in the early stages after the breakup of LB, the evaporation of sessile droplets containing bacteria is primarily controlled by capillary flow. It is believed that both flow regimes influence the deposition and pattern dynamics of bacterial agglomeration in sessile drops. Usually, in pendant drop, migration of bacteria along the interface toward the regions of higher curvature. After detachment of LB, CA of sessile and pendant drops range between 12± 4°, 10± 2°, respectively. The pendant drop leaves a coffee-ring deposit identical to that dried in the sessile mode.

CA of the sessile colloidal drop was detected to be 30± 4°, and the corresponding initial contact diameter ($D_{s0}$) was measured to be ~2 mm. A variable "α" denotes the non-dimensional confinement distance, as $h_{LB}$ is normalized with respect to $D_{lb0}$. $h_{LB}$ was adjusted to a suitable height, and the same confinement distance was maintained throughout the evaporation process. A higher value of α indicated a larger gap between the two surfaces. The present study assesses the dynamics of evaporation by using three values of α (0.11, 0.20, and 0.38), respectively, corresponding to 0.210, 0.370, and 0.450 mm $h_{LB}$. The colloidal LB was left to dry in a controlled environment of ambient temperature conditions of 25°C± 3°C and 45± 3% of the relative humidity. The bacteria (*Salmonella* Typhimurium (STM)) was endogenously tagged with red fluorescent proteins (RFP) to encase the contrasting deposition patterns after evaporation of the LBs ($t_{ev}$ ≈ 10−32 mins), where $t_{ev}$ is the total evaporation time.

## Evaporation dynamics of LB

The deposition patterns after drying of LB primarily rely on evaporation dynamics, wettability, colloidal size and shape, temperature, relative humidity, and confinement distance between the substrates. Therefore, before analyzing the deposition pattern, it would be desirable to understand the evaporation dynamics of colloidal LB. The evaporative phenomena of LB are different from a sessile droplet because of the existence of two independent three-phase CL and both contribute to the overall motion of the interface. Even a slight variation in surface properties could alter the evaporation dynamics [12].

The temporal variation of the non-dimensional CD ($D_{lb}^*$) at the bottom surface is estimated by normalizing the instantaneous CD ($D_{lb}(t_{lb})$) of LB with respect to the corresponding initial CD ($D_{lb0}$) (as shown in Figure 3(a)). The period during which the drop remains in its LB configuration while undergoing evaporation before its separation into individual drops is referred to as $t_{lb}$. The normalized time against the individual $t_{lb}$ represents as $t_{lb}^*$. The diameter decreases with time and shows similar trends for all confinement distances and mediums. The $D_{lb}^*$ at the breakup point also decreases with confinement distance. At a low confinement distance (α = 0.11), the maximum reduction in $D_{lb}$ is observed, and breakup occurs nearly at $D_{lb}^*$ ≈ 0.25. For intermediate confinement distance (α = 0.20), breakup occurs early at $D_{lb}^*$ ≈ 0.4. Since the CL is almost static, hardly any changes are detected in the contact diameter at a higher confinement distance (α = 0.38). LBs exhibit completely different evaporation behaviors with low and high confinement distances, and the evaporation rates mainly depend upon the gap between the plates [12]. In conjunction with the bottom surfaces, CD variations on the top surface also show similar time-dependent behavior. The temporal evolution of the profile of LB displays the triple-line stick-slip behaviour during the evaporation process (refer to Figure S1 in the Supporting Information).

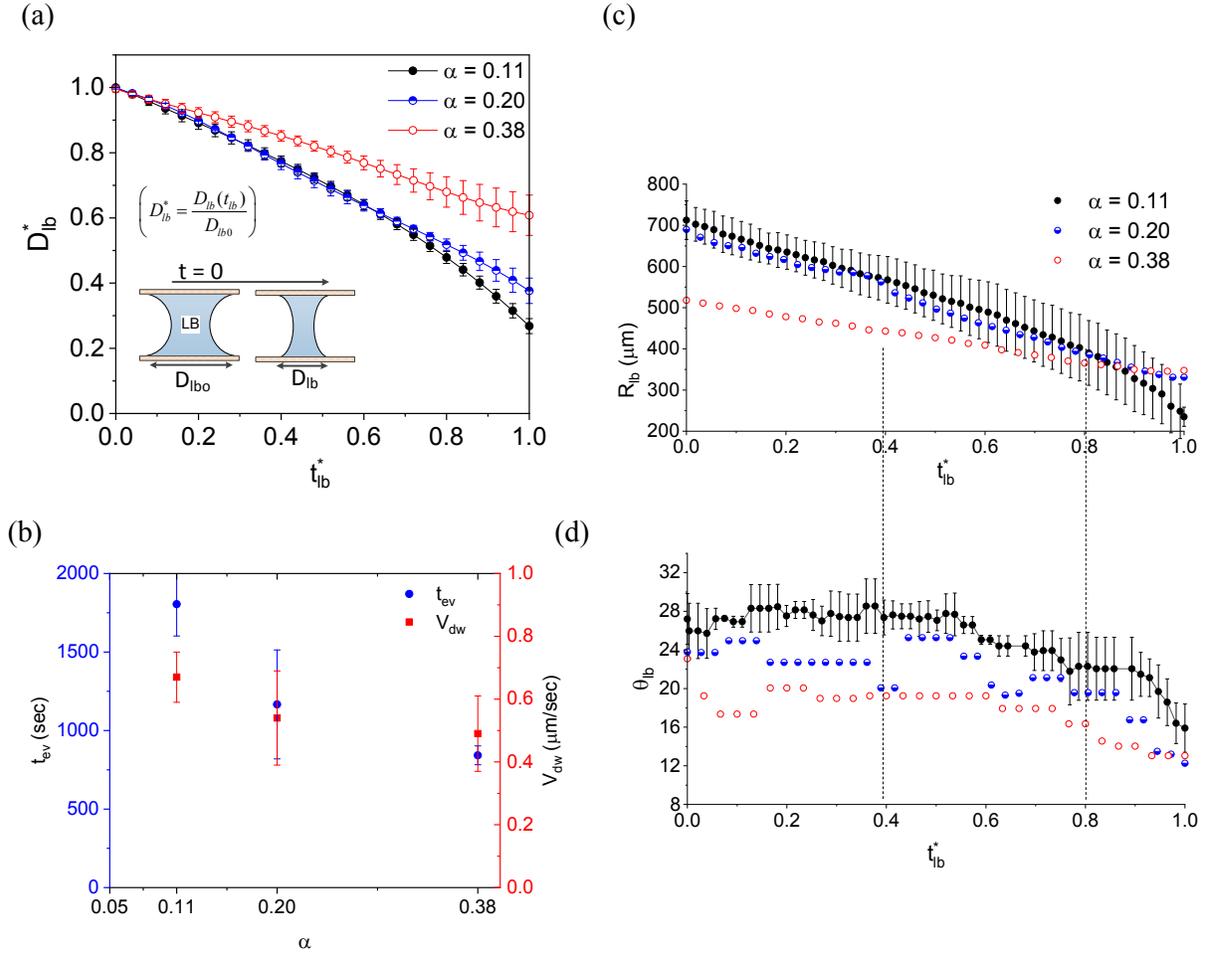

**Figure 3**. For three different confinement distances (α = 0.11, 0.20, and 0.38).
(a) Temporal variation of non-dimensional contact diameter $\left(D_{lb}^* = D_{lb}(t_{lb})/D_{lb0}\right)$ with time $t_{lb}^*$ (normalized against the individual $t_{lb}$). The maximum mean error in estimating $D_{lb}^*$ is within a limit of ±10%.
(b) Evaporation time of LB ($t_{lb}$) and dewetting velocity (estimated from experimental observations) as a function of confinement distance.
(c) Dynamics of contact radius ($R_{lb}$) and (d) contact angle ($\theta_{lb}$) of LB against non-dimensional time $t_{lb}^*$ (normalized against the individual $t_{lb}$) for three different confinement distances (α = 0.11, 0.20, and 0.38).

The transient variations of contact radius and CAs (bottom of the curvature of corresponding LBs) of different confinement distances are represented in Figure 3(c-d). During stick mode, CA and contact radius decrease, while during slip mode, CA constant and contact radius decrease more (jump appears). At a low gap between plates more stick-slip result multiple rings pattern however at a high gap less stick-slip results pinned–single ring pattern formed (refer to Figure 1 c). In the present study, all the gaps, with time, $\theta_{lb}$ exhibited an oscillation trend up to a critical time threshold $\left(t_{lb}^* \sim 0.8\right)$. The magnitude of the initial CAs decreases as the gap between the surfaces increases, whereas the CAs are higher at the low value of α. This is because, for a given volume of liquid, the reconfiguration of colloidal droplets for LBs with higher confinement distance leads to reduced initial CDs with lower slopes of the liquid meniscus at the three-phase contact point.

Confinement distance significantly affects the evaporation time ($t_{lb}$) and dewetting velocity, as shown in Figure 3(b), and the evaporation time of LB represents the considerable difference in time scales between the formation of the stable LB and its breakup. With an increase in the confinement distance between the plate surfaces, $t_{lb}$ of the LB decreases significantly. A smaller CA produces a larger evaporation flux at the plates [12], and the related phenomena are evident in the higher gap between the plates at smaller CAs. Hence, the evaporation rate slows down with a decrease in confinement distance. Experimental results also showed that two times increase in α results in ~60% reduction of $t_{lb}$ (Figure 3(b)).

From the understanding developed so far, the CA and the CL play a vital role in evaporation. CL follows a continuous and discontinuous stick-slip motion at various stages of evaporation. The dewetting velocity ($V_{dw}$) is estimated by measuring the average displacement of CL over the entire timescale ($t_{lb}$) during the period of drop volume remaining in an LB configuration. The calculated velocity ($V_{dw}$) decreases monotonically as the confinement distance is increased (~ $O(10^{-7})$ m/s) (refer to Figure 3(c)). It would appear that this result is supported by the characteristics shown in Figure 3, since a maximum reduction in CD was observed for the case of the highest dewetting velocity. There is a possibility that a higher dewetting velocity is caused by the stick-slip motion of the LB drop, while a low $V_{dw}$ is indicative of CL that is nearly stationary.

The $V_{dw}$ can be used to quantify the shear stress on the cell. A bacterium experiences the maximum stress when it adheres to a solid surface. Consider now a linear velocity profile across the cross-section of the bacterium, the shear stress τ, scales as [54,55]

$$\tau \sim \mu \frac{V_{dw}}{d} \qquad (1)$$

Where $\mu$, $V_{dw}$, and $d$ are the viscosity of fluid (for milli-Q $1.1 \times 10^{-3}$ Pa s), dewetting velocity, and cell diameter (~ 1 μm), respectively. It yields shear stress of about 7 μPa, 5 μPa, and 4.5 μPa on bacterium for the confinement distance 0.11, 0.20, and 0.38 for milli-Q, respectively. It shows that shear stress decreases with confinement distance; however, the magnitude of this decrease is very small.

**3.2. Theoretical analysis on evaporation of liquid bridge**

In this work, an analytical solution of Upadhyay & Bhardwaj (2021) for the evaporation rate of a drop forming an LB between two flat surfaces, is extended for a hyperbolic curve of an LB that is symmetric about the midplane. Moreover, the predicted profiles are then compared to experimentally determined profiles.

We obtain the simultaneous time evolution of surface area $A(t)$ and volume $V(t)$ of the LB during the whole drying process until break-up from captured images. The captured images are automatically processed using image processing tools from Matlab. The shape of the LB ($r(z)$ and $r'(z) = \partial r / \partial z$) are computed on each image. More specifically, the LB was segmented into subparts, and a circular revolution was employed to assess all the sub volumes (refer to Figure 4). This method accurately evaluated the LB volume and surface area [56]. The volume was subsequently obtained by summing up these individual subsections. From those measurements, we deduce the surface area $A$ and volume $V$ by

$$A = \int_{-h/2}^{h/2} 2\pi r \sqrt{1 + r'^2}\, dz \text{ and } V = \int_{-h/2}^{h/2} \pi r^2\, dz \qquad (2)$$

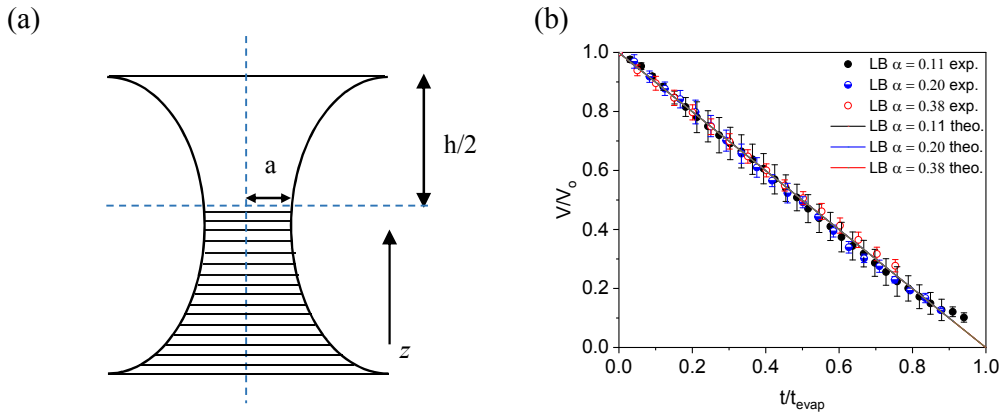

**Figure 4.** (a) Schematic view of LB for volume evolution. The notation $a$ is the neck radius, and $h$ is the height of the LB.
(b) Volume ratio $V/V_0$ versus normalized time $t/t_{evap}$ of LB configuration at different gaps for the same initial volume. The solid line represents the theoretical model, while the experimental data is depicted using markers. The notations exp. and theo. refer to experimental and theoretical volume regression scales, respectively.

In the present study, we estimate the evaporation flux ($J_{th}$) and instantaneous mass loss (evaporation rate) of LB by the given relation [12]

$$J_{th} = \frac{D(1-RH)c_{sat}}{r \ln\left(\frac{D_\infty}{2r}\right)} \qquad (3)$$

$$\frac{dm_{LB}}{dt} = A_s \frac{D(1-RH)c_{sat}}{r \ln\left(\frac{D_\infty}{2r}\right)} \qquad (4)$$

Where, $m_{LB}$ is the mass of the LB, $A_s$ is the surface area, $r$ is the radius of geometry, and $D_\infty$ is the distance between interface and ambient (refer to Fig. 1b). The above relation is based on the diffusion of liquid vapor in surrounding air and is governed by the Laplace equation for the liquid−vapor concentration (c,kgm$^{−3}$). The above equation was derived by performing an order of magnitude analysis by solving the Laplace equation for the liquid−vapor concentration in a cylindrical coordinate system and assuming the interface to be cylindrical. In the absence of external convection, the diffusion of liquid vapor in the surrounding air is governed by the Laplace equation. Moreover, a negligible variation of the vapor concentration in the vertical direction is assumed. The experiments were arranged in a manner where the liquid was squeezed between the plates at a distance from the free edge, resulting in a small ratio of $D_{wet}/D_\infty$ (less than 1). Consequently, we can neglect the influence of external convection. In the present study, the small Bond number $\left(Bo = (\rho g h^2/\gamma) \sim 10^{-2}\right)$ indicates that gravity is insignificant for our experiments.

When a squeezed droplet evaporates under steady state, isothermal diffusion-controlled conditions, its mass decreases with time as follows,

$$\dot{m}_{LB} = \frac{dm_{LB}}{dt} = \frac{2\pi h D(1-RH)c_{sat}}{\ln\left(\frac{D_\infty}{2r_c}\right)} \qquad (5)$$

Here, $r_c \left(=\sqrt{V/\pi h}\right)$ represents the instantaneous radius of the equivalent cylinder. Initial volume ($V_0$) of the droplet taken from the experiment. The observations could be corroborated by calculating the theoretical instantaneous mass evaporation.

Next, we focus our analysis to compute the evaporation time of LB. First, we start from the evaporation of a squeezed droplet, which forms an effective cylinder in still air, and then extend it to a more complicated geometry of liquid bodies, e.g., LB. To evaluate the instantaneous volume of the droplet by considering the cylinder, Eqn. 5 is solved. The Eqn. (5) shows that increasing $h$ leads to increased mass loss and, hence, faster evaporation. Also, as $r_c$ decreases with evaporation, the instantaneous mass loss reduces. Now, this mass evaporation rate could be used to advance the solution in time with a step size *dt*. The time step was taken as 1 ms after carrying out a time-step convergence analysis. To do so, the net mass of the LB at the next time step was calculated by subtracting $\dot{m}dt$ from the mass ($m_t$) at time *t*. The updated mass was then used to find the volume ($V_t = m_t/\rho$). The process was repeated until a finite known value of $V_t$. The total time spent in the process could be used to estimate the evaporation time for a particular configuration. As mentioned above, LB allows it to evaporate at a fixed height. Therefore, with time, only the radius of the cylinder is updated.

Figure 4(b) shows the comparison of the experimental normalized volume ratio $V/V_0$ as a function of the normalized time coordinates $t/t_{evap}$ with theoretical prediction for LB configuration for $\alpha$ = 0.11, 0.20 and 0.38. It is also noted form Fig. 4(b) that evaporation linearly depends on time, as observed by others [56].

In the present study, LB formed between hydrophilic substrates where the initial contact angle ranges from 32° to 25°. Similar geometry has also been observed for LBs forming between two surfaces as the geometry reflects the surface with a minimal area. Similar geometry has also been observed when a soap film is stretched between two axial rings, and the resultant geometry is found to be a catenoid [12]. However, in the present work, the profile of an LB between two surfaces is fitted by a hyperbolic cosine function. A typical equation for a hyperbolic cosine function is given by

$$r_{lb}(z) = a \cosh\left(\frac{cz}{a}\right) \quad (6)$$

Here, $r_{lb}$ is the relative radial coordinate at height *z* from the center line, *a* is the geometric parameter defining the curvature of the shape of the neck radius, and *c* is the scaling constant of stretching. For *c* = 1, the above equation represents a standard catenoid. In the present work, variable *z* is multiplied by a constant *c* whose value is greater than 1. If the wettability of both plates is the same, then the bridge wets both surfaces symmetrically. The volume of the LB is expressed as

$$V = \int_{-h/2}^{h/2} \pi \left( a \cosh\left(\frac{cz}{a}\right) \right)^2 dz \quad (7)$$

For the constant height (*h*), variables *a* and *c* can define the volume. One can also now estimate the surface area $A_s$ of the interface between any two vertical coordinates –*h*/2 and *h*/2 using the following relation.

$$A_s = 2\pi \int_{-h/2}^{h/2} r_{lb}(z) \sqrt{1 + \left[r'_{lb}(z)\right]^2} dz$$

$$= 2\pi \int_{-h/2}^{h/2} a \cosh\left(\frac{cz}{a}\right) \sqrt{1 + \left[c \sinh\left(\frac{cz}{a}\right)\right]^2} \, dz \tag{8}$$

The evaporated mass can be estimated by substituting $r_{lb}(z)$ in Eqn. (5), in a similar way as volume calculated from experimental images.

$$\Delta \dot{m} = 2\pi D(1-RH)c_{sat}Q(z) \tag{9}$$

The term $Q(z)$ can be estimated by summation of the subsections ($\Delta z = 1$ µm) over the height of the LB.

$$Q(z) = \sum_{i=1}^{n} \left[\frac{\Delta z}{\ln\left(\frac{D_\infty}{2a\cosh(cz_i/a)}\right)}\right] \tag{10}$$

Now, to calculate the evaporated mass from Eq. (9), both neck radius and stretching scale need to be determined from the profile of LB. Variation of neck radius ($a$) with time taken from the experiments. However, stretching scale $c$ can be estimated by the volume relation (Eqn. (7)). The solution method is similar for both hyperbolic geometry of Eqn. (9) and to the cylindrical geometry of Eqn. (5).

Figure (5) shows the comparison of the experimental evaporation characteristics with the theoretical predicted by Eqn. (5) for cylindrical geometry and Eqn. (9) for hyperbolic geometry while keeping same initial volume. From Figure 5(a), we can observe that the evolution of $V/V_0$ predicted by hyperbolic geometry is much closer to the experiment value than that of cylindrical geometry. This directly reflects the role of surface area and justifies the catenoid assumption. Moreover, from Figure 5(b), we can observe that the evaporation time predicted by hyperbolic geometry is much closer to the experiment value.

The variation in the instantaneous volume ratios obtained from our model with the experimental observations might be attributed to the assumption of continuous slip. The highest confinement distance ($\alpha = 0.38$) provides the lowest CD reduction. As the CL appears to be almost static, we observe very little change in CD. In contrast, the evaporation behavior of squeezed LBs differs substantially from that of higher $\alpha$ cases. In such instances, the stick-slip motion controls the mass transfer mechanism, resulting in a notable decrease in CD. Since the influence of temperature and relative humidity on evaporation is clear from the model prediction, we assumed constant temperature and relative humidity in all cases, i.e. 298 K and 50%, respectively.

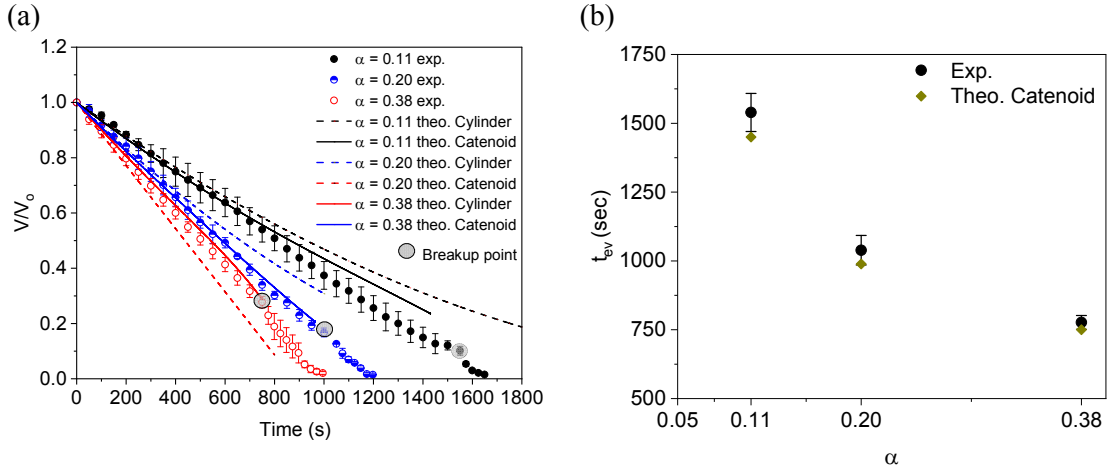

**Figure 5.** A comparison between the experimental results and theoretical model based on cylindrical and hyperbolic approaches at three different confinement heights α = 0.11, 0.20, and 0.38 for the same initial volume of milli-Q. (a) variation in the volume ratio ($V/V_0$) over time in seconds, and (b) evaporation time. The notations exp. and theo. refer to experimental and theoretical volume regression scales, respectively.

Due to the symmetry of LBs about both the central axis and the midplane, each quadrant profile reveals all the properties of the entire bridge, such as the separation distance or height between two surfaces and the volume of the liquid. Figure 6 compares the calculated LB profiles (solid line) for different time intervals with those obtained from experiments (symbols) for three confinement distances. The agreement between the measured and predicted profiles is excellent for the higher gap. The discrepancies come from the uncertainties in the relative humidity and the volume measurement. In addition, the role of stick slip motion is important, as the contact angle variation will have an effect on the bridge profile. The minor offsets in the lower gap (α = 0.11) profiles shown in Figure 6(d) are consistent, which may distort the profile, while the stick-slip motion still governs the motion. In addition to the surface tension, the gap of the LB will also affect the curvature of LBs. The bridge shape is believed to remain stable at every moment in time, allowing stretching to be addressed in a quasi-static manner.

A theoretical model based on diffusion-driven evaporation is used to characterize the timescale of the evaporation of the LB as well as its profile. The model's predictions align well with the current experimental findings for hyperbolic cosine function compared to cylindrical assumption.

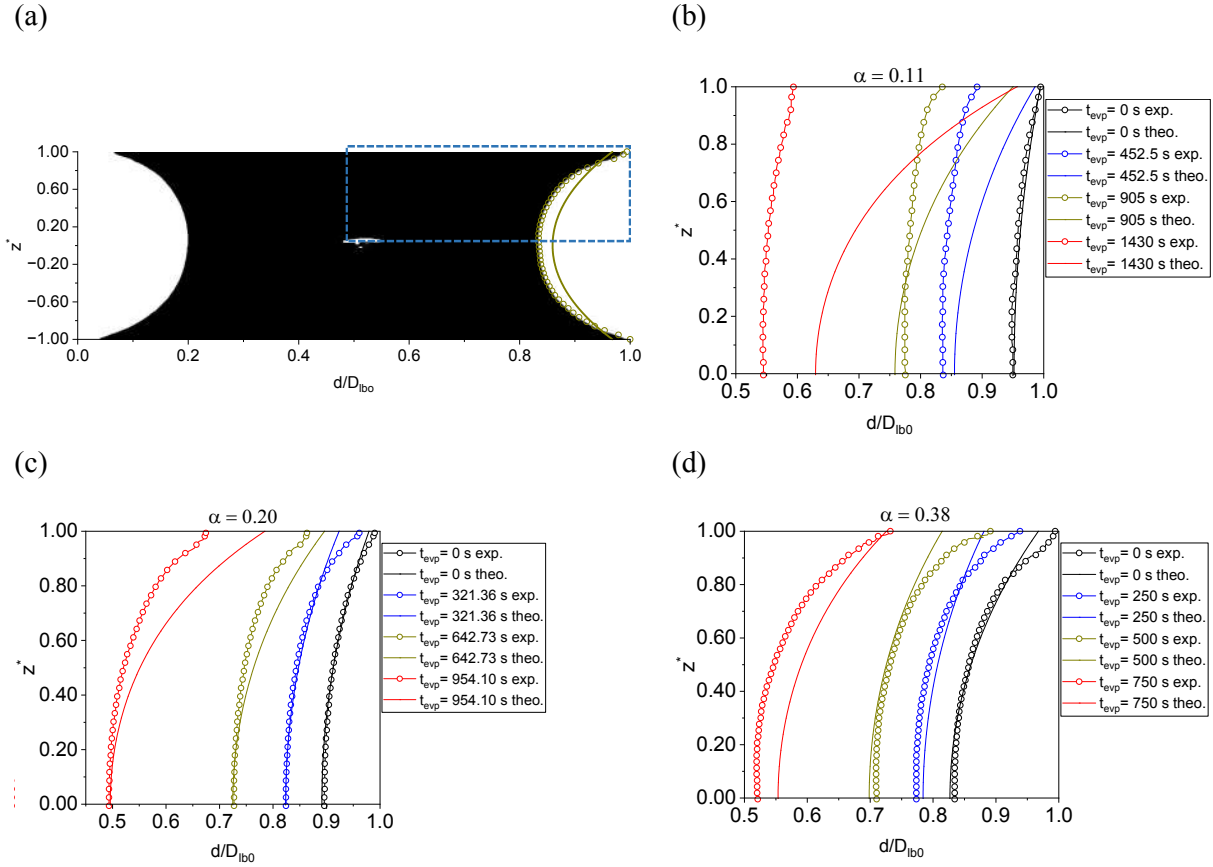

**Figure 6.** Image of profiles of LB with volume V~ 0.4 ±0.1 μL between two identical glass surfaces for milli-Q. (a) shows the entire profiles of LBs between glass surfaces. (b), (c) and (d) are comparisons between predicted and experimental profiles of the LBs of the upper right quadrants for α = 0.11, 0.20, and 0.38, respectively. The symbols are the experimentally measured points, while the solid curves are the predicted profile. The notations exp. and theo. refer to experimental and theoretical volume regression scales, respectively.

### 3.3. Deposition patterns

After understanding the evaporation dynamics of bacterial LB, we will subsequently focus on the deposition pattern on the top and bottom of the glass substrate. Bacteria aggregation after evaporation of colloidal droplets has been more complex than inert particles and ascertained different pattern formations. However, it should be noted that, unlike a sessile droplet, there are two independent triple-phase contact lines for a LB, and both will contribute to the overall motion of the interface. The evaporation pattern of the bacterial LB has not been reported till now to the best of the author's knowledge.

To determine the effect of evaporation mode on the final deposit pattern, we compared LB precipitates and sessile precipitates maintained with the same volume of bacterial solution and identical evaporation conditions. The sequence of the deposited patterns on the top surface plate after the evaporation of LB can be divided into three stages, as illustrated in Figure 7. These stages are the initial stable LB ($t^*≈0$), the receding of LB ($0<t^*<1$), and certainly after the detachment of LB (sessile and pendant drop). In this regard, the present study considered a typical case of bacterial solution (STM) with milli-Q for characterization and behavioral sequence of the bacteria in the deposited patterns. From the discussed sequential process, one can observe the dense deposition of the bacteria with multiple layers at the CL, zigzag (honeycomb or network) pattern while receding, and thick deposition at edges during evaporation of droplet are presented in Figure 7 for the top surface plate.

A stick-slip motion has been observed in the receding of the LB, in which the contact line is repeatedly pinned and depinned (refer to Figure 7). Whereas in initial and after detachment only particles are pinned at the contact line. In Figure 7, symbols *a* and *b* denote the typical slip points. The "Stick" behavior at the contact line is due to the accumulation of particles along the LB edge because of the evaporation-driven flow. The "Slip" behavior, on the other hand, is caused by the contact angle receding as the LB evaporates, as it has reached its static value. The droplet's equilibrium contact angle is quickly established when the contact line separates, as the line slips to its subsequent equilibrium position and adheres there. The stick-slip motion of the contact line is caused by the competition between friction and surface tension forces until the drop is completely evaporated, leaving multi-rings of LBs [57–59]. In the present study, stick-slip motion has been observed in evaporating LBs at all three confinement distances (refer to Figure 7). The effect of the contact angle on the stick-slip motion is not observed here [60].

In the case of higher confinement distance, the receding of the LB for a finite distance is observed during the evaporation process i.e. nearly constant CL, and after the breakup, liquid volume retracts the CL of the initial stage (Figure 7c). Therefore, the deposits in the second stage (during the receding) likely overlapped with sessile and pendant drop deposits that allowed the bacteria to accumulate at the CL, resulting in a central coffee ring.

For lower confinement distances, the receding distance of LB is higher, and the reduction in CD is also evident from the plot data (Figures 7 (a), and 7(b)) as compared to the higher value of α. Due to the continuous depletion of solvent (water) in such cases, LB exhibits a stick-slip motion, influencing the evaporation of a higher volume of liquid in view of lower evaporative driven flow, as discussed earlier. Therefore, after the breakup, the sessile and pendant drop volume reduces and does not cover the CL of the initial stage. As a result, the deposit patterns of bacteria during the initial receding of LB and deposition of the droplet are clearly distinguished in the microscopic images, as shown in Figures 7(a) and 7(b).

For lower and intermediate confinement distances, the bacterial deposited at CL and the formation of the outer ring can be seen at initial stable LB as well as the thick line structure of the inner ring after the breakup. Following the disintegration of LB, both sessile and pendant drops result in evaporation through capillary driven flow conditions. It is observed that the lower and intermediate confinement distance, the alteration of CL, and the receding of LB would affect the final deposit compared to the deposit by the sessile drop. However, when the confinement distance was higher, the bacteria of the LB deposited in a pattern resembling a coffee ring, similar to the deposition of a sessile drop without LB (as shown in Figure 7d).

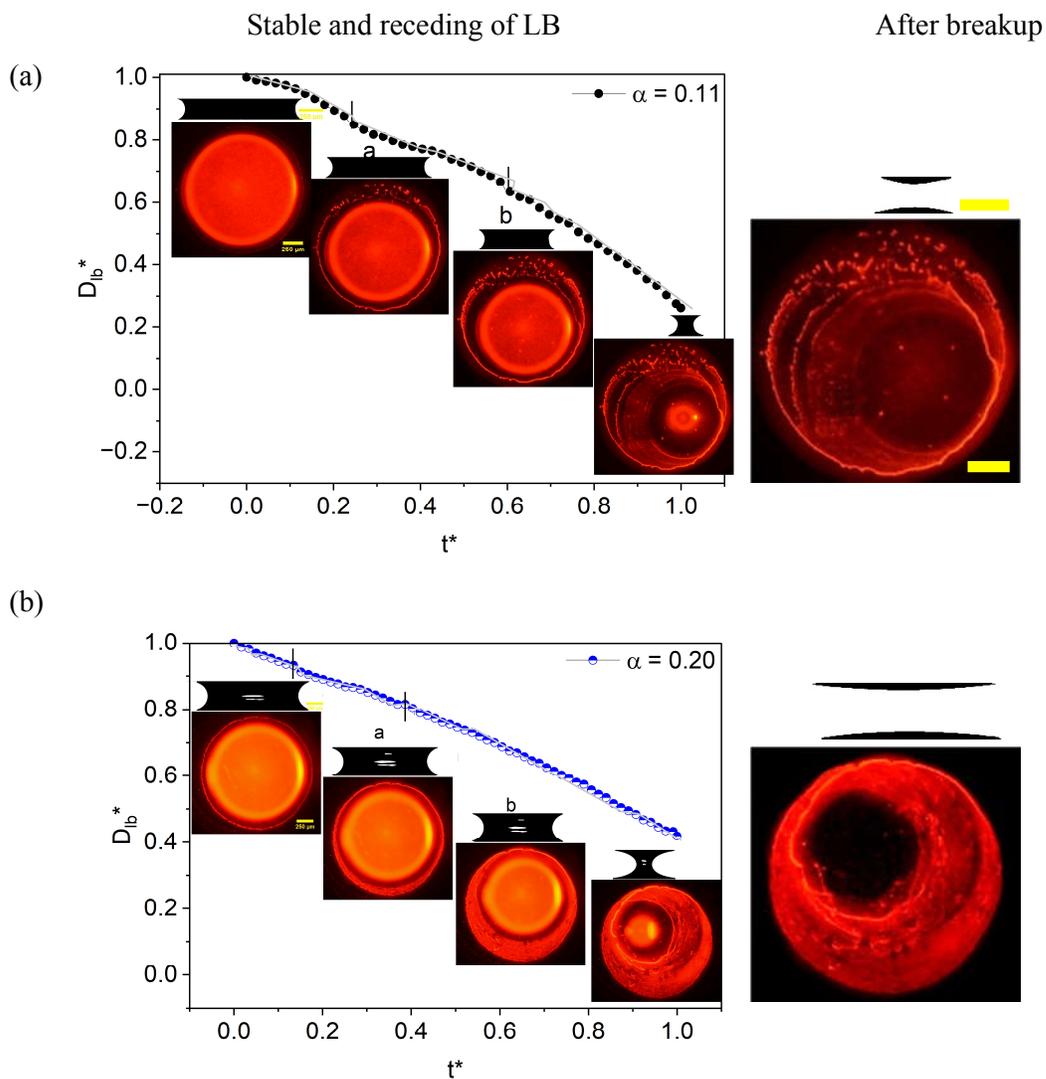

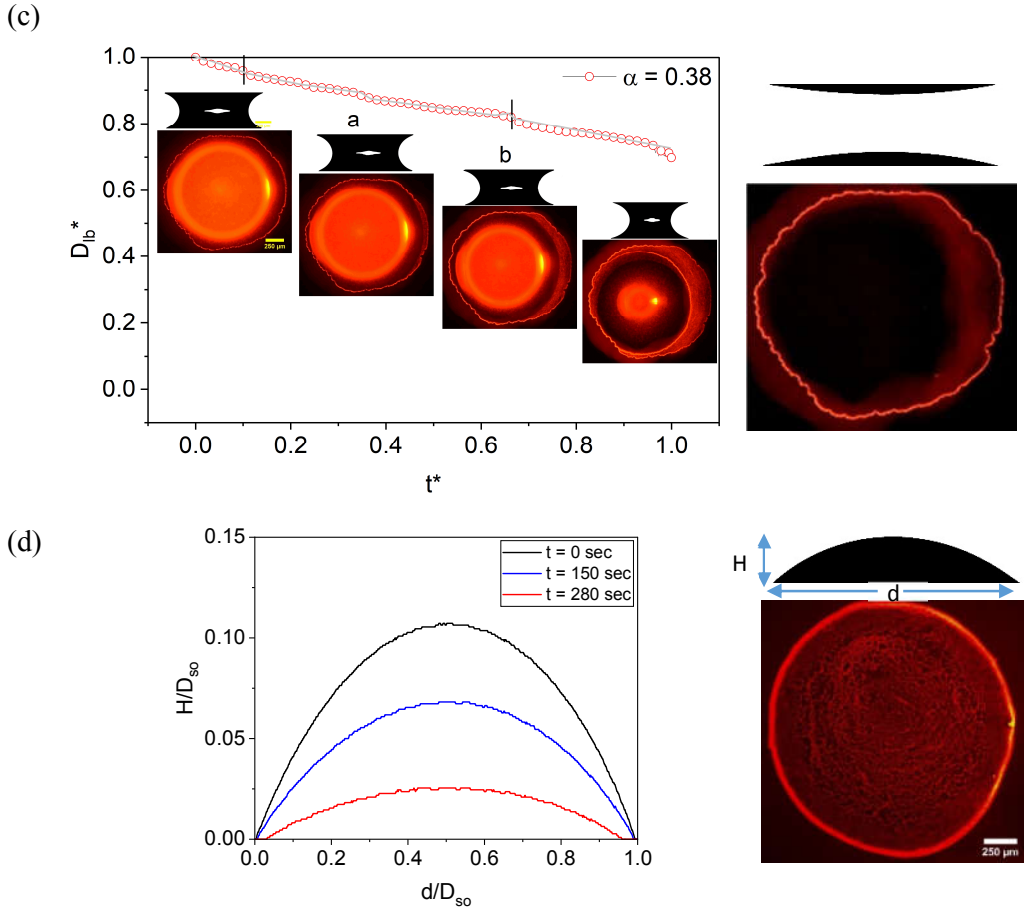

**Figure 7.** Normalized LB contact diameter, $D_{lb}/D_{lb0}$, as a function of the normalized time ($t^* = t/t_{lb}$) measured from side-view images with solid line shows the stick and slip motions of the contact line, (a) $\alpha = 0.11$, (b) $\alpha = 0.20$, and (c) $\alpha = 0.38$. Front side images with deposited pattern on the top plate are shown at different stages of evaporation. (d) Normalized sessile CD $H/D_{so}$ as a function of the $d/D_{so}$. Notations H and d refer height and diameter sessile droplet. The scale bar denotes 250 $\mu$m.

After evaporation of the bacteria-laden (STM) droplet, the deposition morphology in the vicinity of the LB periphery was recorded via microscopy, presented in Figure 8, with three different confinement distances. Generally, the evaporation dynamics of the LB depending on the variation of the bridge height, and different deposited patterns were ascertained from the captured images.

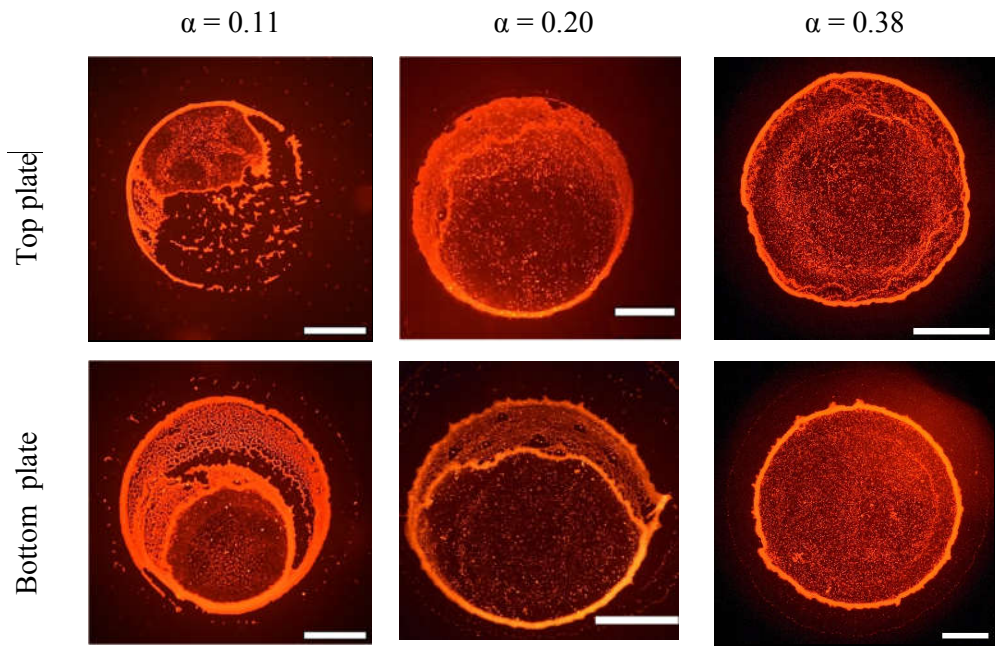

**Figure 8.** Varying final deposit patterns of bacterial-laden (STM) LB. (a) Alteration of confinement distance (α ranges from 0.11 to 0.38) leads to variation in precipitation patterns observed at the top and bottom surfaces. All the fluorescence images are acquired at 20× magnification. Scale bars, represented by the white line, are equal to 0.5 mm.

In this context, at lower and intermediate confinement distances (α ≈ 0.11 and 0.38), the thin film evaporation is observed (discussed in the next section), which leads to multiple rings and network pattern formation in both top and bottom surfaces in milli-Q solution. As discussed above stick-slip motion is responsible for multiple ring formation. However, the network pattern is due to thin film instability (refer to next section). It is clear from Figure 8 that at a low confinement distance, we have not observed any scallop shell like pattern [14], which indicates that the behavior of bacteria is not identical with non-interacting particles. In the case of the higher confinement distances (α ≈ 0.38), a coffee ring pattern structure can be seen.

The quantitative analysis of the LB precipitates for neutral solutions reveals denser edge deposition around the droplet periphery. This leads to very dense deposits of bacteria near the CL. As discussed above, at low confinement distance, deposit patterns are affected by the continuous sticking and slipping motion of the CL. The bacteria interact widely in the same way as sticky particles bond with one another when they come into contact [26]. It is reported that the similarity of the deposition pattern on the top and bottom substrate when the colloidal size is less than 1.1 μm [12]. In the present work, the bacterium size is in the range of 1 μm [61]; even then, the deposition pattern at the top slide is not completely identical to the bottom pattern.

Furthermore, to understand bacteria activeness on the deposit patterns, similar experiments were performed on LB with the application of dead bacteria. It is reported that active and dead bacteria produce different patterns upon evaporation, and bacterial deposition cannot depend upon the nutrients [29,31]. The experiments for LB evaporation of the dead STM and PAO1 bacteria with nutrient-neutral medium (milli-Q) were carried out. However, nearly identical deposited patterns were seen on the surfaces from the microscopic images for the dead bacteria (STM) with milli-Q, as presented in Figure 9. The deposition patterns of dead bacteria after disintegration of the LB (sessile and pendant drops) form disc-like and spot-like shape dried patterns that exhibit different flow interactions in the evaporation of sessile and pendant drops through the capillary flow and Marangoni flow conditions [62]. However, bacterial morphology is discussed in more detail in the further subsections.

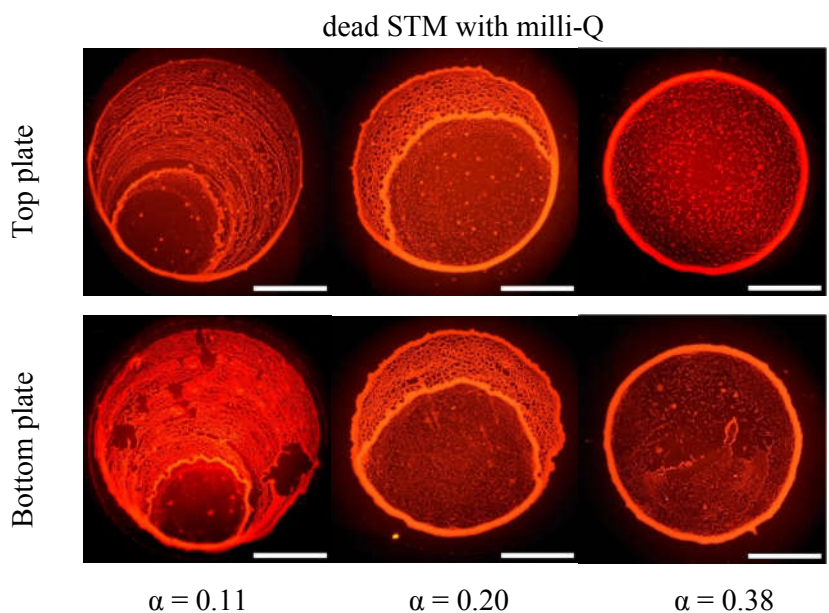

**Figure 9.** Final drying precipitates of dead STM in milli-Q medium at three different confinement distances ($\alpha$ = 0.11, 0.20, and 0.38). Scale bars, represented by the white line, are equal to 0.5 mm.

To further investigate the deposition of dried LB, PAO1 was used to observe bacterial behavior. Deposition patterns of live PAO1 in the milli-Q are quite identical to STM. The supplementary Figure S2 shows the overall picture of bacterial deposition patterns of live PAO1 and dead PAO1 with milli-Q.

### 3.4. Thin film interferometry

The deposited patterns discussed earlier have seen the honeycomb weathering (network) for the LB configuration in low confinement distance. As mentioned previously, the honeycomb pattern forms due to the slower evaporation rate of LB and the receding of thin film between the two plate surfaces at the low confinement distance. The variation of the thin film along the time intervals after the establishment of LB is recorded with the reflection interference microscopy technique. The alignment of the experimental apparatus and the necessary equipment are shown in Figure 2.

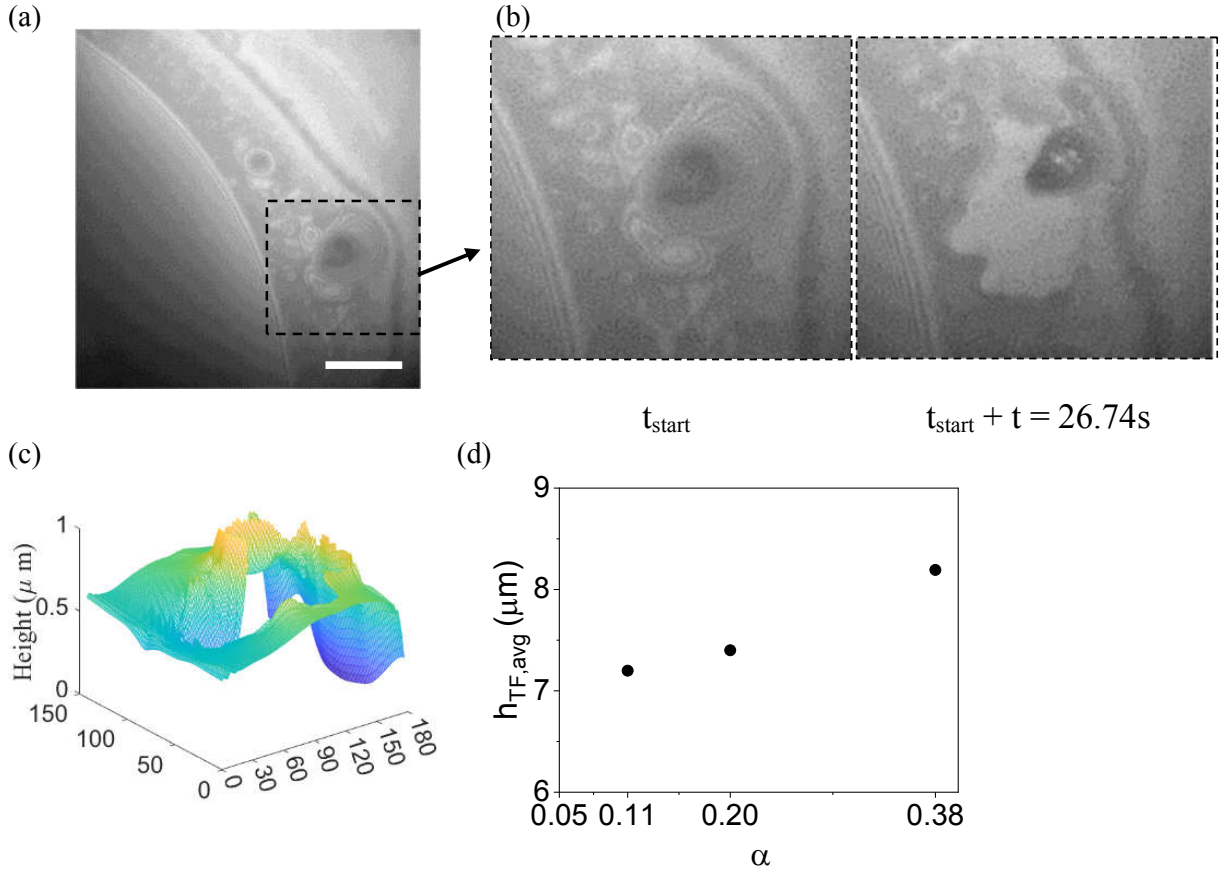

**Figure 10.** Interference images were captured at 50 fps for STM laden LB for milli-Q at low confinement distance α = 0.11. The scale bar is 100 μm. (a,b) insights of dewetting of thin film layer (c) variation of the thickness profile along the dewetting phenomena of thin film instability at the CL of LB (d) Thin-film measurement profiles for the different fluid mediums of bacterial laden LBs. $h_{TF,avg}$ denotes the averaged thin-film height.

The reflected light source from the bottom plate surface passes through the beam splitter, and the change in the phase difference of the light allows to form of an interference pattern on the screen. The interference patterns were used for the measurement of thin film profiles for the various fluid mediums by employing image-processing frequency guided algorithms. The fringe patterns associated with water laden bacterial LBs are shown in Figure 10(a). At the initial time instant, the existence of the liquid film from the contact line to the three-point of LB is observed in both cases. The fringe patterns were observed at the contact line of LB at the initial time instant, and the presence of dense fringe patterns was seen in the later time instant that induced possible variation in film thickness, where the liquid film is about to dewet or rupture due to evaporation of thin film as shown in Figure 10. The measurement of two-dimensional profiles of the thin film at the breakup region was demonstrated in Figure 10(c), which represents the movement of the liquid film towards the edges while the film is de-wetting and the significant variation of the height profile along the edges (in microns ≈1.5 µm) is evident from the obtained fringe patterns while receding of the LB. As de-pining of the LB initiates from the edge, patch formation starts closer to the edge. It can be seen in Figures 10 (a) and (b) that patch formation proceeds towards the central region as the film thickness decreases. In addition to the patch that is forming near the LB edge, another patch is forming adjacent to the LB center as well. As a result, a bacterial island is formed when the growing fronts from two patches merge. During the evaporation, a number of dry patches form and will repeat these merging to form the network patterns.

The varying thin film height at the CL of the LB for different fluid mediums was shown in Figure 10 (d) for the considered confinement distances of the LB. The increase in film thickness is evident with an increase in the confinement distance for all base fluid mediums. The distribution of the local thickness profile and contact angle live STM with milli-Q is shown in supplementary Figure S3.

### 3.5. Bacterial morphology

From the discussions of the deposited patterns in the earlier section, the inertial and capillary forces are found to have a significant role in altering bacterial morphology. Capillary force is inversely proportional to the gap between the plates [10]. The topographic images of the bacterial (live STM) cell wall at three confinement distances ($\alpha = 0.11$, 0.20, and 0.38) were obtained using atomic force microscopy (AFM) in nutrient-neutral (milli-Q), shown in Figure 11. The left-hand column shows a 20 µm square area picture of the deposit edge, whereas the right-hand column indicates only a magnified enlarged 3D image in a 5 µm square area.

At the edge of the LB, bacteria are arranged as a major axis parallel to the CL (refer to Figure 11) in the milli-Q base medium. Bacteria tend to attach to one another near their poles as the continuous capillary flow motion during droplet evaporation influences the bacteria to form a confinement line structure (Figure 11). The thickness of the bacterial deposit increases with $\alpha$ (the related details were explicitly presented in the supporting file Figure S4). The variation of the thickness with increased $\alpha$ is due to the alteration of the contact angle of LB, the dewetting velocity, and the increase in the gradient of concentration flux along the surface of LB. As seen from Figure 3 (c), dewetting velocity is lower at a higher value of $\alpha$, which leads to dense and multiple layers depositions of bacteria near the CL.

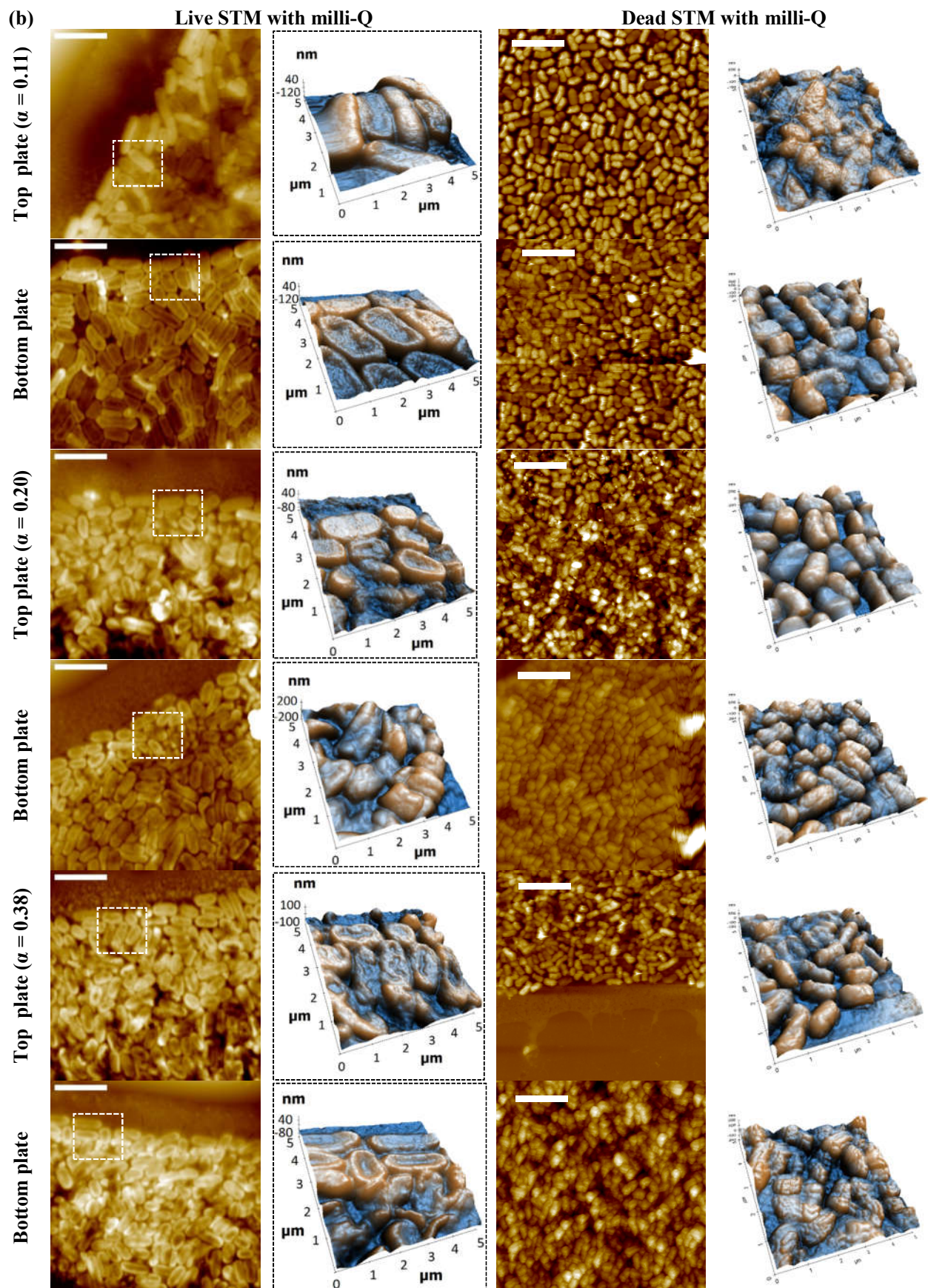

**Figure 11**. (a) Typical edge and network region from the deposited pattern
(b) AFM image shows deposited bacterial morphology and self-assembly at the edge of the top and bottom surface of the LB for nutrient-neutral (milli-Q)base fluid solution of dried STM (WT) at three different confinement distances α = 0.11, 0.20, and 0.38 in a 20 μm square area and enlarged 3D image in a 5 μm square area. The scale bar is 5 μm.

As the CL starts depinning, forms the thin film layer of liquid induces the thin-film instability as the thickness of liquid layer decreases. The presence of bacteria in a thin microscopic liquid layer (Figure 10) perturbs the variation in thickness, resulting in multiple dewetting processes of the liquid layer to form the network patterns [52].

The bacterial morphology of STM is similar to PAO1 irrespective of bacteria status (live or dead). The supplementary Figure S5 shows the overall picture of bacterial deposition at the edge of live and dead PAO1 with nutrient-neutral (milli-Q) as a base fluid medium.

Moreover, one can observe that the bacteria get deposited at the CL through the capillary flow movement. The rupture of the thin-film at multiple locations along the CL leads to stretch deposition of the bacteria. With the disintegration of thin film near the end of evaporation, the bacterium becomes points of accumulation for more bacterial deposition on the substrate in small clusters over the surface. Furthermore, the bacteria deposited as a monolayer at the network pattern (Figure 12), and the edge bacteria are arranged with a major axis and self-assembled due to the continuation of rupture phenomena of the thin-film instability from the contact line to the center of the droplet.

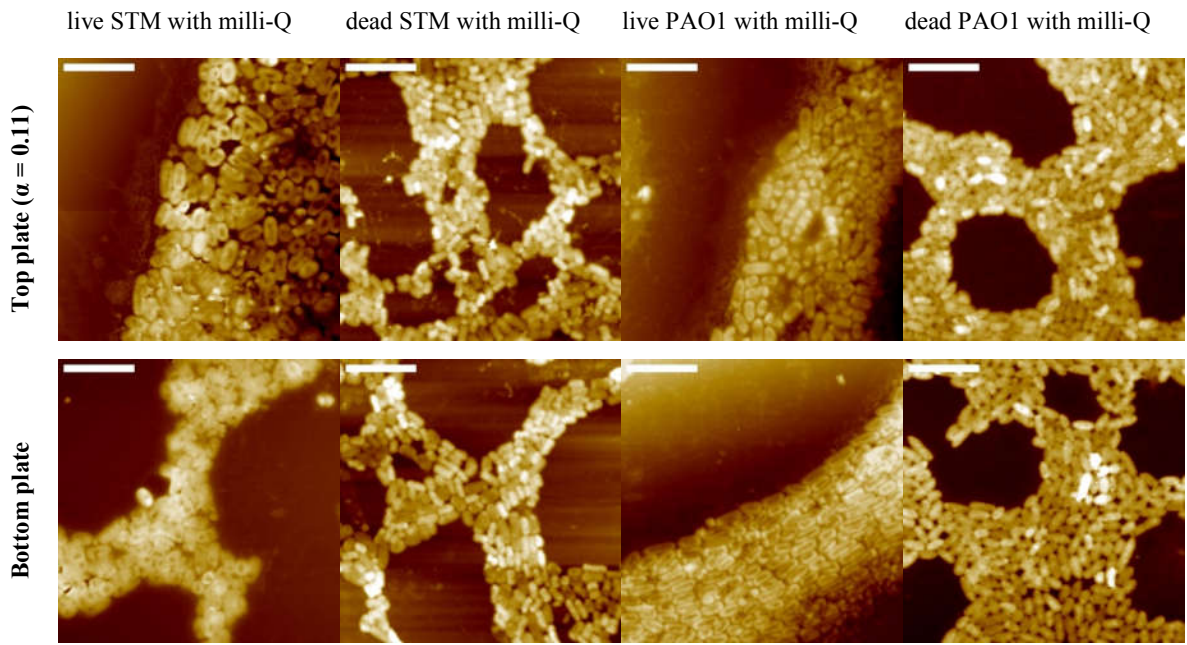

**Figure 12**. AFM image shows deposited bacterial morphology and self-assembly at the network pattern for milli-Q for α= 0.11, at the top and bottom surface of the LB, STM, and PAO1 for both live and dead. The scale bar is 5 μm.

Further, we investigate the size distributions of bacterial populations at the edge of LB. We measure the flat surface area by interface detection of all cells at the edge of the top and bottom surface from the AFM images (Figure 13(a)). Comparing the area at the top and bottom of the deposit of STM and PAO1 (live and dead) in the milli-Q medium through probability density function in Figure 13(a-f). The cells are capable of adjusting their size and shape in response to mechanical forces in their environment [63]. The bacterial cell wall is subjected to capillary adhesion force as well as evaporative stress, which does not allow the bacteria to gain its initial cell wall shape.

The aerial size of the bacterial cell increases with decreasing the gap between the plates for both live and dead STM and PAO1 with milli-Q medium (Figure 13(a)-13(f)) as a result of more adhesion force. Bacterium bulging can be seen for milli-Q solution from the zoom image of 5 µm in Figure 13. The bacterial height of *Salmonella* when subjected to desiccation stress and the mechanical stress increased from 74 ($\alpha = 0.11$) to 79 nm ($\alpha = 0.38$) for the bottom plate in milli-Q medium (Figure S4). Dead bacteria have rigid cell structure that restricts their deformation [25], for both STM and PAO1. Therefore, dead cells have a lower area compared to live cells, and the variation is noticeable from the obtained experimental data, as shown in Figure 13.

### 3.6. Mechanism of bacterial self-assembly in thin film

The evaporation dynamics of LB can be divided into two stages: before breakup (evaporation of stable LB) and after the breakup of LB (evaporation of sessile and pendant drop). For the evaporation dynamics of a stable LB, it is necessary to understand the fundamentals of thin-film evaporation processes (schematically shown in Figure 14a). An evaporating meniscus can be divided into three hypothetical regions based on the force acting on it: non-evaporating film region, transition region (thin film evaporation region), and bulk fluid region [64,65]. Among the three regimes, the transition region is the most significant for the final deposition of bacteria into the surface, where the liquid film thickness in this regime decreases below the bacterial length scale [66].

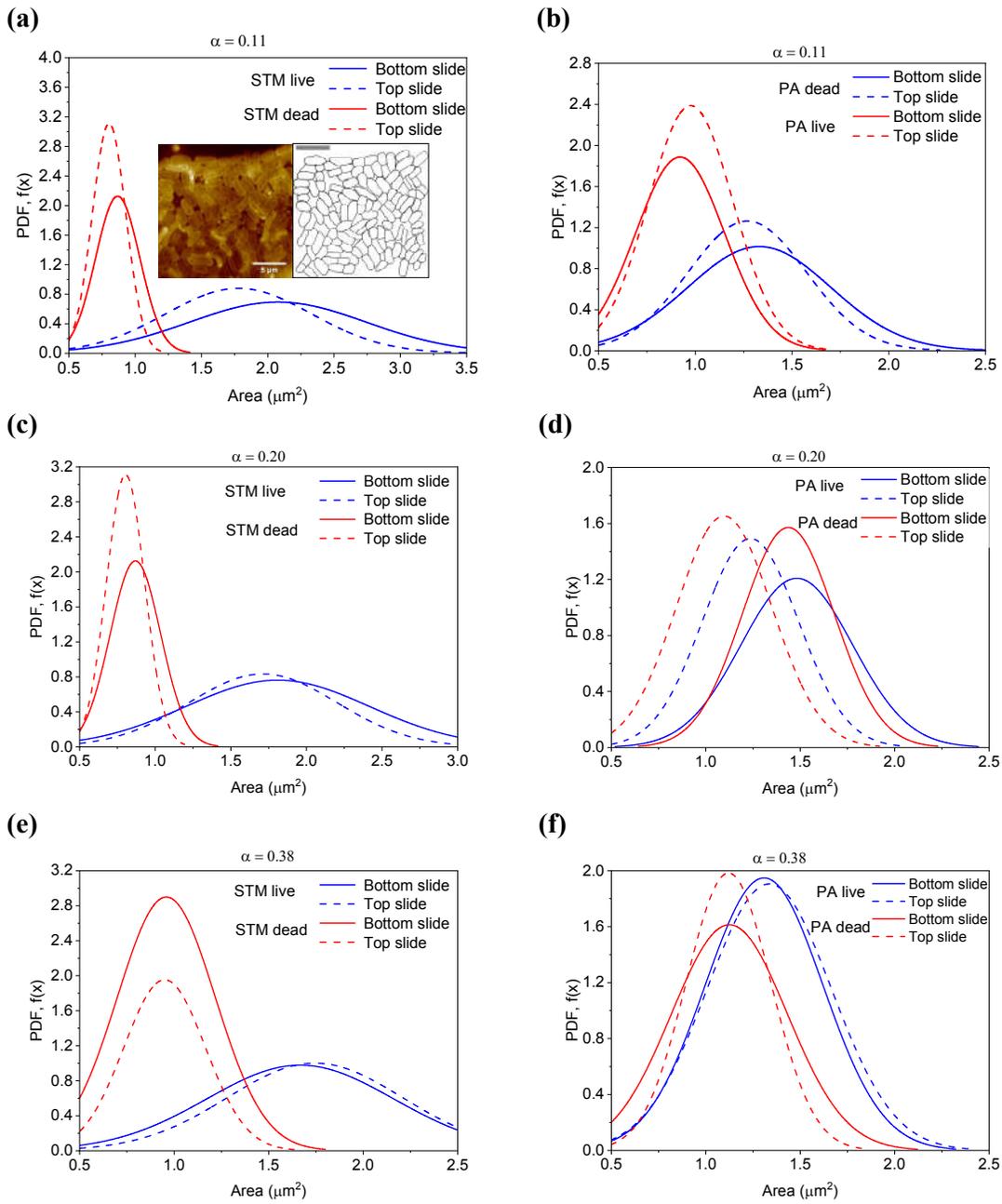

**Figure 13.** Bacterial aerial size distribution at edge of the LB, (a)AFM image and interface detection of cell, Plot comparing the area at the edge of the deposit through Probability density function (PDF), for milli-Q (STM live and dead) at (a) α = 0.11, (c) α = 0.20, (e) α = 0.38, PDF for milli-Q (PAO1 live and dead) at (b) α = 0.11, (d) α = 0.20, (f) α = 0.38.

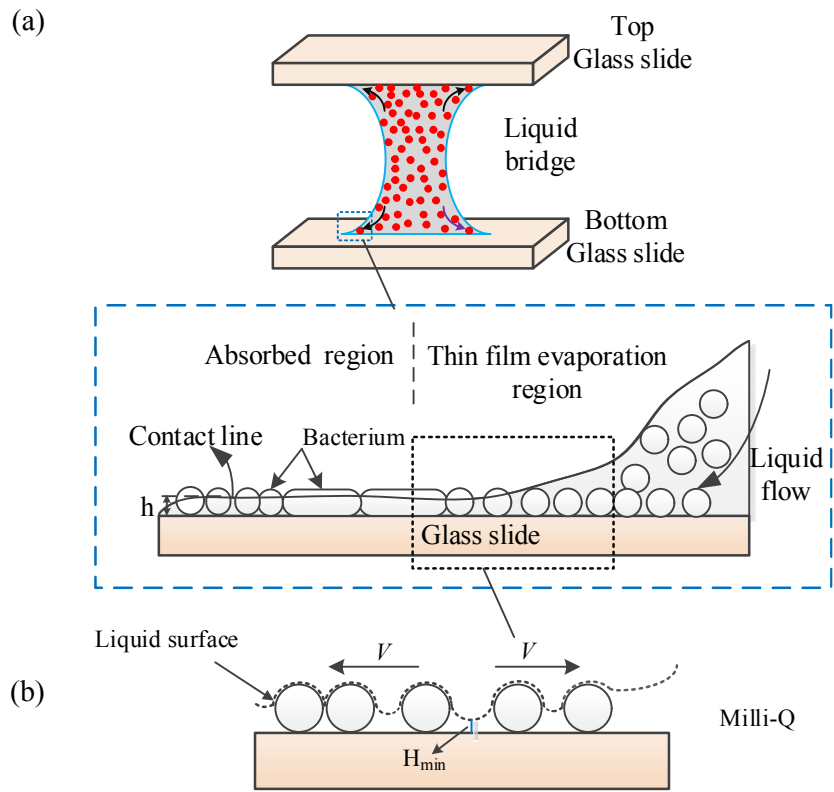

**Figure 14**. Mechanistic insight on bacterial self-assembly (forming network pattern) in thin film region.

At the initial stage of evaporation, bacteria accumulate at the three-phase CL of LB compliance to form a coffee ring like pattern due to capillary flow, usually for all the fluid mediums. Therefore, the monolayer of bacteria would start to form when the CL starts receding, and the thickness of the liquid layer is equal to or lesser than the size of the bacterium. However, at the later stage of evaporation (after forming a coffee ring at the outer edge of LB), the deposition patterns may depend upon evaporation dynamics, concentration, the attraction among bacteria (self-assembly), and the collective dynamics of bacteria. A mutual and surface attachment of bacteria can make the deposition process complex with unpredictable dynamics.

In this context, for the fluid medium of milli-Q water, few bacteria in the bulk fluid region advance in an arbitrary direction. At the same time, the existence of capillary flow drags bacteria toward the LB edge, which may cause the bacteria to agglomerate and stick to the substrate at random locations as thin film height reaches the bacteria length scale. As a result of the active motion of the bacteria, there are slight deflections in the trajectory, and orientations can be seen at the outer edge of the LB. As time progresses, the thin film capillary instability and the presence of bacteria stuck onto the substrate become points of accumulation for more bacteria, eventually producing small clusters like honeycomb weathering. The formation of network patterns at low and intermediate confinement distances is the result of mechanisms of particle (bacterium) assembly and liquid film dewetting at the final stage of the evaporation of colloidal droplets [67]. Thin film rupturing is related to transition thin film evaporation region where bacteria are confined between liquid-air interface and substrate (Figure 14b). An indentation of the capillary bridge forms between the two or more cells of the bacteria. As the liquid between the bacteria evaporates, the local liquid film is thinning and forming a concave surface. When the liquid film is accomplished to the minimum thickness, it results in unbalanced capillary forces and, hence, the motion of particles; this might be one of the underlying factors driving dry patch formation. In the case of higher confinement distance ($\alpha$ = 0.38), the height of the thin film evaporation region is higher (as shown in Figure 10) to get capillary flow drag bacteria toward the LB edge. Therefore, dry patches are absent in higher confinement distances.

In the second stage of evaporation (after the breakup of LB), a sessile drop is formed in the bottom and a pendent drop in the top slide. The capillary flow is responsible for forming a coffee ring at the outer edge of the drop. The drying pattern for sessile and pendant drops is nearly identical for contact angle below 45° and particle size below 3 μm [68]. The deposition patterns may depend upon the attraction among bacteria and the collective dynamics of bacteria. Bacteria can attach to each other and to surfaces, thereby making the deposition process complex and unpredictable.

## 4. CONCLUSIONS

The present study experimentally and theoretically investigated the evaporation of the liquid bridge for the bacterial suspensions that have been confined between two parallel plate surfaces. Squeezing of the liquid droplet between the two plates modifies the curvature of the LBs results in a significant alteration in the evaporation flux across the liquid–vapor interface, and as well as the plausible effect of the mechanical stress induced by the evaporation dynamics demonstrates the regulation of the bacterial size based on the neutral mediums.

First, we have studied the evaporation dynamics of the LB for different gaps between the plates experimentally and theoretically. A theoretical model based on the analytical solution of the simplified Laplace equation was used to estimate the evaporation flux mass-loss rate. The liquid−vapor interface is approximated using a cylindrical and hyperbolic cosine function geometry, and diffusion-limited evaporation is assumed. The model predictions align well with the current experimental results.

At a lower confinement distance of LB, multiple base patterns (coffee rings and network) were observed on the bottom substrate for the fluid water to the live and dead bacteria of STM and PAO1. In case of higher confinement distance, a single coffee ring pattern in both top and bottom plates is similar to the deposit of the sessile case. The data acquired from the evaporation dynamics of the different confinement distances and the residual deposits on the surfaces of the bacterial LBs depend mainly on the stick-slip motion of the CL coupled with the evaporation dynamics and tendency of the bacteria to form biofilms by sticking to one another. The variation in the thin film instability due to stick-slip motion at the low confinement distances was assessed from the interferometry images, and the dry patterns in network patterns were evident from the microscopic images, marking the one-to-one comparison of the results.

AFM imaging depicts the variation in size of the bacteria, which increases with a decrease in the confinement distance. Moreover, in the present study, the aerial size of live bacteria was greater than that of the dead bacteria. The confinement spacing and fluid medium greatly influence the morphology of the patterns and the spatial distribution of bacteria. Despite the fact that the study has been conducted using *Salmonella Typhimurium* and *Pseudomonas aeruginosa*, we are inclined to believe that similar results may be observed with other types of bacteria as well.

Analytical solutions for predicting evaporation time can be derived based on mathematical models that describe the physical processes involved. However, the reported model for estimating the evaporation time of the LB does not take into account the effect of the stick-slip motion. As an extension of the present study, it is important to investigate the effect of stick-slip motion at the low confinement height of LB.

## ASSOCIATED CONTENT

**Supporting Information**

The supporting information is available free of charge. The time-varying outline of LB (milli-Q, $\alpha = 0.38$), fluorescence microscopy images as well as AFM images of live and dead PAO1 with milli-Q for different confinement distances, thin liquid layer interferometry results milli-Q, Bacterial deposit height live STM with milli-Q solution (PDF)

Video S1: Typical shadowgraphy data ($\alpha = 0.38$) depicting evaporation of LB (AVI)

## AUTHOR INFORMATION

**Corresponding Author**

**Saptarshi Basu** − *Department of Mechanical Engineering, Indian Institute of Science, Bangalore 560012 Karnataka state, India; orcid.org/0000-0002-9652-9966; Email: sbasu@ iisc.ac.in*

## Authors

**Kush Kumar Dewangan -** *Department of Mechanical Engineering, Indian Institute of Science, Bangalore 560012, Karnataka, India*

**Srinivas Rao S -** *Department of Mechanical Engineering, Indian Institute of Science, Bangalore 560012, Karnataka, India*

**Durbar Roy-** *Department of Mechanical Engineering, Indian Institute of Science, Bangalore 560012, Karnataka, India*

**Atish Roy Chowdhury-** *Department of Microbiology & Cell Biology, Indian Institute of Science, Bangalore 560012, Karnataka, India*

**Dipshikha Chakravortty-***Department of Microbiology & Cell Biology, Indian Institute of Science, Bangalore 560012, Karnataka, India; Email: dipa@iisc.ac.in*

## Present Addresses

† **Durbar Roy-** International Centre for Theoretical Sciences, Tata Institute of Fundamental Research, Bangalore 560089, Karnataka, India

## Author Contribution

‡ S.R.S., D.R. contributed equally to the work.

## Author Contribution

Conceptualization: S.B. and D.C.; Methodology: S.B., D.C., K.K.D., S.R.S., D.R., and A.R.C.; Investigation: K.K.D., S.R.S., D.R., and S.R.S.; Visualization: K.K.D., S.R.S., D.R., and A.R.C.; Supervision: S.B. and D.C.; Writing - original draft: K.K.D., D.R., S.R.S., and A.R.C.; Editing and revision: S.B., D.C., K.K.D., D.R., A.R.C., and S.R.S.

## Funding Sources

The authors disclosed receipt of the following financial support for the research, authorship, and/or publication of this article: This work was supported by the Science and Engineering Research Board (PDF/2021/ 002796).

## Notes

The authors declare no competing financial interest. All relevant data are within the paper and its supporting information files. All materials and additional data are available from the corresponding author upon request.

ACKNOWLEDGMENT

SB acknowledges financial support from the Science and Engineering Research Board (SERB), Pratt & Whitney Chair Professorship and SERB-SUPRA (Scientific and Useful Profound Research

Advancement) project number SERB/F/10572/2021-2022. The work is supported by the Science and Engineering Research Board (SERB). The first author (KKD) acknowledges the support of the SERB National Post-Doctoral Fellowship (PDF/2021/002796) awarded to him. The authors would like to thank the Indian Institute of Science (IISc), for providing the facility for AFM.

ABBREVIATIONS

Kush Kumar Dewangan, KKD; Srinivas Rao S, SRS; Durbar Roy, DR ; Atish Roy Chowdhury, ARC; Dipshikha Chakravortty, DC; Saptarshi Basu, SB